\documentclass[oneside,reqno,a4paper,12pt]{amsart}

\usepackage{latexsym}
\usepackage{euscript}
\usepackage{epsfig}
\date{\today}
 \large\normalsize
\newcommand{\be}{\begin{equation}}
\newcommand{\ee}{\end{equation}}

\def\ie{{\it i.e.}}
\def\LEP2{{LEPII}}
\def\npb#1#2#3{    {\it Nucl. Phys. }{\bf B #1} (19#2) #3}
\def\npbps#1#2#3{    {\it Nucl. Phys. }(Proc. Suppl.){\bf B #1} (19#2) #3}
\def\plb#1#2#3{    {\it Phys. Lett. }{\bf B #1} (19#2) #3}
\def\prd#1#2#3{    {\it Phys. Rev. }{\bf D #1} (19#2) #3}

\def\prl#1#2#3{    {\it Phys. Rev. Lett. }{\bf #1} (19#2) #3}
\def\ptp#1#2#3{    {\it Prog. Theor. Phys. }{\bf #1} (19#2) #3}

\begin{document}
\begin{titlepage}
\begin{flushright}
BA-99-51
\end{flushright}
\vskip 1.5cm
\title[]{{\small Low energy consequences from supersymmetric
models\\ \vskip 0.2cm with left-right symmetry}} \maketitle
\begin{center}
\textsc{Shaaban Khalil$^{1,2}$ and Qaisar Shafi$^1$} \\
\vspace*{1cm} \small{\textit{$^1$Bartol Research Institute,
University of Delaware Newark, DE 19716.}} \\ \vspace*{2mm}
\small{\textit{$^2$Ain Shams University, Faculty of Science, Cairo
11566, Egypt.}} \\
\end{center}
\vspace*{2cm}
\begin{center}
ABSTRACT
\end{center}
\vspace*{8mm}
\begin{quotation}
We consider several low energy consequences arising from a class
of supersymmetric models based on the gauge groups $SU(2)_L\times
SU(2)_R \times U(1)_{B-L}$ and $SU(4)_C\times SU(2)_L \times
SU(2)_R$ in which the gauge hierarchy and $\mu$ problems have
been resolved. There are important constraints on the MSSM
parameters $\tan \beta (\simeq m_t/m_b)$, $B$ and $\mu$, and we
discuss how they are reconciled with radiative electroweak
breaking. We also consider the ensuing sparticle and Higgs
spectroscopy, as well as the decays $b\rightarrow s \gamma$ and
$\mu \rightarrow e \gamma$. The latter process may be amenable to
experimental tests through an order of magnitude increase in
sensitivity.
\end{quotation}
\setcounter{page}{1}
\end{titlepage}
\section{Introduction}
In a couple of recent papers~\cite{shafi,dvali}, the minimal
supersymmetric standard model (MSSM) arose from the low energy
limit of a special class supersymmetric models based on
$SU(3)_C\times SU(2)_L \times SU(2)_R \times
U(1)_{B-L}$~\cite{left} and $SU(4)_C\times SU(2)_L \times
SU(2)_R$~\cite{salam}. By imposing a suitable $R$-symmetry,
$U(1)_R$, which contains the unbroken $Z_2$ `matter' parity of
MSSM, it was shown that this class of models has some interesting
`low energy' consequences. For instance, the magnitude of the
supersymmetric $\mu$-term of MSSM gets related to the
supersymmetry (SUSY) breaking scale $m_{3/2}$. The $B \mu$ term of
MSSM is also generated and found to be of order $m^2_{3/2}$,
while the MSSM parameter $\tan \beta \simeq m_t/m_b$, where $m_t$
and $m_b$ are the top and bottom quark masses respectively. The
apparent stability of the proton ($\tau_p
> 10^{32} -10^{33}$ yrs.) is understood to be a consequence of an
`accidental' global $U(1)_B$ symmetry. The $SU(4)\times SU(2)_L
\times SU(2)_R$ model also suggest the existence of `heavy'
charge $\pm e/6$ (colored) and $\pm e/2$ (color singlet) states.
\vskip 0.3cm Motivated by these results we propose to investigate
additional `low energy' implications of these left-right
symmetric models. In particular, we would like to focus on the
important issues of radiative electroweak (EW) breaking,
sparticle and Higgs spectroscopy, composition of the lightest
supersymmetric particle (LSP), and implications of the radiative
processes $b\rightarrow s \gamma$ and $\mu \rightarrow e \gamma$.
Since $\tan \beta (\simeq m_t/m_b)$ is large and the parameter
$B\mu$ is also constrained, the requirement of radiative EW
breaking turns out to be non-trivial. In particular,
non-universal soft SUSY breaking terms and some deviation from
the minimal K\"ahler potential must be considered. The
requirement that SUSY correction to the bottom (b) quark mass
should not be excessive ($\leq 20 \%$) imposes additional
constraints on the parameters of the model.
 \vskip 0.3cm
The plan of the paper is as follows. In the next section (2), we
briefly describe the underlying left-right symmetric models, the
mechanism for resolving the $\mu$-problem, and the origin of the
$B\mu$ term. We also discuss deviations from Refs.~\cite{shafi,
dvali} needed to obtain a $B$ term that is consistent with
radiative EW breaking. In section (3) the EW symmetry breaking is
discussed in detail, while constraints arising from the b-quark
mass are taken up in section (4). Section (5) deals with the
ensuing SUSY spectrum as well the composition of the LSP. The
corresponding Higgs spectroscopy is briefly considered in section
(6). Sections (7) and (8) focus on the radiative processes
$b\rightarrow s \gamma$ and $\mu \rightarrow e \gamma$
respectively. Our conclusions are summarized in section (9).
\vskip 0.3cm
\section{The Model}
\vskip 0.35cm For definiteness, we will take the underlying
symmetry group to be $G=SU(4)_C\times SU(2)_L \times SU(2)_R$ and
follow the notation used in Ref.~\cite{shafi}. The breaking of
$G$ at the GUT scale ($M_{GUT}$) to $SU(3)_C\times SU(2)_L \times
U(1)_Y$ is achived by introducing non-zero vacuum expectation
values (VEVs) for the Higgs superfields $H$ and $\bar{H}$, which
transform under $G$ as:
\begin{eqnarray}
H &=& (4,1,2), \nonumber\\ \bar{H}&=&(\bar 4,1,\bar 2).
\end{eqnarray}
The MSSM Higgs doublets are contained in the representation $h$
of $G$, where \begin{equation} h = (1,2,2).\end{equation} A color
sextet superfield $D=(6,1,1)$ is also included to ensure that the
`low energy' particle sector coincides with that of MSSM.
Finally, the quarks and leptons belong to the
$(4,2,1)_i+(\bar{4},1,2)_i$ representations of $G$, where
$i=1,2,3$ denotes the generation index. \vskip 0.3cm The
superpotential of the minimal $G$ model is given by (after
imposing a suitable $U(1)_R$ symmetry)~\cite{shafi,dvali},
\begin{eqnarray}
 W & = & S[\kappa (\bar{H} H  - M^2) +  \lambda  h^2]
+ \lambda_HDHH + {\lambda}_{\bar{H}} D \bar{H} \bar{H} \nonumber
\\ & + &  \lambda_{33} \bar{F}_3 F_3 h +\lambda_{ij} \bar{F}_iF_j
h\frac{(\bar{H} H)^n}{M_P^{2n}} + \lambda_{\nu ij} \frac{\bar{F}_i
\bar{F}_j H H}{M_P} , \label{susypot}
\end{eqnarray}
where $S$ denotes a gauge singlet superfield, the parameters
$\kappa, \lambda$ and $M$ can be taken to be real and positive,
and $h^2$ denotes the unique bilinear invariant $\epsilon^{ab}
h^{(1)}_a h^{(2)}_b$. Also, $M_P (\simeq 2.4 \times 10^{18}$ GeV)
denotes the `reduced' Planck mass.  The Higgs fields develop VEVs,
$\vert \langle H \rangle \vert = \vert \langle \bar{H} \rangle
\vert \simeq M$, which lead to the symmetry breaking
\begin{equation}
SU(4)_C\times SU(2)_L\times SU(2)_R\to SU(3)_C\times SU(2)_L\times
U(1)_Y. \label{422:321}
\end{equation}
Note that supersymmetry is unbroken at this stage. The inclusion
of soft SUSY breaking terms will induce an expectation value
(proportional to $m_{3/2}$), namely \begin{equation} \langle
S\rangle = - \frac{A_{\kappa} - A_1}{2 \kappa^2}.
\end{equation}
Here and throughout, as is customary, the scalar components of
the superfields are denoted by the same symbols as the
corresponding superfields. $A_{\kappa}$ and $A_1$ denote the
coefficients of the soft trilinear and linear terms that contain
$S$. This means that the $\lambda \langle S\rangle h^2 $ term in
eq.~(\ref{susypot}) provides an effective MSSM $\mu$ parameter of
the correct order of magnitude. With $A_{\kappa}= \kappa A
m_{3/2}$, $A_1 = \kappa (A-2) m_{3/2}$ (minimal supergravity),
\begin{equation}
\mu = - \frac{\lambda}{\kappa} m_{3/2}. \label{mu1}
\end{equation}
Furthermore, the bilinear term is given by
\begin{equation}
B = 2 m_{3/2}.
\end{equation}
\vskip 0.3cm This model implies Yukawa unification for the third
family (see eq.(\ref{susypot})),  which leads to a large top mass
$m_{t} > 165$ GeV and $\tan \beta \sim m_{t}/m_{b}$~\cite{shafi2}.
The first and second family Yukawa couplings, as well as mixings,
eventually must be generated by non-renormalisable operators
and/or the inclusion of additional states. We will not address
this important issue here. \vskip 0.3cm Since $\tan \beta \gg 1$,
it is necessary to employ non-universal soft SUSY breaking to
satisfy the EW breaking conditions~\cite{kobayashi2}. However, it
turns out that, even with non-universal soft SUSY breaking, the
condition \begin{equation}\sin 2\beta = \frac{2 B \mu }{m_{H_1}^2
+ m_{H_2}^2 + 2 \mu^2}\end{equation} cannot be satisfied with the
value of $B$ at $M_{GUT}$ of order $2 m_{3/2}$. We can modify the
value of $B$ as follows. Consider the relevant superpotential
terms, namely
\begin{equation}
 \delta W =  S[\kappa (\bar{H} H  - M^2) +  \lambda  h^2] ,
\end{equation}
which leads to the potential
\begin{eqnarray}
\delta V &=& \vert \kappa (\bar{H} H  - M^2)+ \lambda h^2 \vert^2
+ \vert S \vert^2 (\kappa^2 \vert \bar{H}\vert^2 +\kappa^2 \vert
H\vert^2 + \lambda \vert h \vert^2) \nonumber\\
     &+& m_{3/2} (\vert S\vert^2 +
\vert \bar{H} \vert^2 + \vert H \vert^2 + \vert h\vert^2 )+
(A_{\kappa} S \bar{H} H + A_{\lambda} S h^2 - A_1 SM^2
+\mathrm{h.c}).
\end{eqnarray}
We now depart from the minimal K\"ahler potential considered in
Ref.~\cite{dvali} by assuming non-universality between the
trilinear couplings $A_{\lambda}$ and $A_{\kappa}$, namely we
assume \begin{equation} A_{\kappa} = \kappa A m_{3/2} ,
\end{equation} \begin{equation} A_{\lambda} = \lambda A'
m_{3/2} , \end{equation} but keep the assumption
\begin{equation}A_1 = \kappa (A-2) m_{3/2}.\end{equation} Then, we still have
$\langle S \rangle = -m_{3/2}/{\kappa}$,  and $\mu =
-\frac{\lambda}{\kappa} m_{3/2}.$ \vskip 0.3cm The bilinear
coupling $B\mu$ is given by \begin{equation} B \mu = \lambda F_S^*
+ A_{\lambda} S .\end{equation} Since \begin{equation} F_S^* =
-\frac{1}{\kappa} (\kappa^2 \vert S \vert^2 + m_{3/2}^2 ) -
\frac{1}{\kappa} A_{\kappa} S , \end{equation} we find that
\begin{equation} B\mu = -2 \frac{\lambda}{\kappa} m_{3/2}^2 +
\frac{\lambda}{\kappa} (A-A')m_{3/2}^2 , \end{equation} so that
\begin{equation} B = \left ( 2 -(A-A') \right) m_{3/2}
.\end{equation} For $A > A'$, we can have $B < m_{3/2}$, which is
needed to realize the EW breaking scenario in the large $\tan
\beta$ case. \vskip 0.3cm
\section{Electroweak Symmetry breaking}
\vskip 0.3cm The phenomenological aspects of models with large
$\tan \beta$ can be quite different from those with small $\tan
\beta$ values. In particular, radiative EW symmetry breaking is
an important issue. This has been discussed under the assumption
of universal soft SUSY breaking parameters in Ref.~\cite{ann,
carena}. In the large $\tan \beta$ scenario the mass squared
parameters for the down (up) sector Higgs $H_1$ ($H_2$) run from
the higher energy scale $M_{GUT}$ to the weak scale $M_Z$ in very
similar ways if these masses are universal at $M_{GUT}$. This is
not conducive for successful EW symmetry breaking, especially in
view of the above constraints on $\mu$ and $B$. Requiring
non-universality at $M_{GUT}$ such as
\begin{equation} m_{H_1}^2
> m_{H_2}^2 ,\end{equation} turns out to be favorable for symmetry breaking
with large $\tan \beta$. Also the trilinear coupling should be
larger than the gaugino masses. Large values of the $A$ parameter
are crucial to reduce the value of $B$ during the running from
$M_{GUT}$ to $M_Z$. Also, radiative breaking requires
non-universality among the gaugino masses. As we will show in the
next section, the supersymmetric correction to the bottom quark
mass constrains the gluino mass to not be very heavy and
therefore implies a constraint on $M_3$. The running of $B$
imposes a constraint on $M_2$, while $M_1$ is essentially
unconstrained. However, at the weak scale it turns out that in
all cases we have $M_1 < M_2 < M_3$. \vskip 0.3cm We will assume
the following boundary conditions on the soft scalar masses at
$M_{GUT}$:
\begin{eqnarray} m_{H_1}^2 &=& m_Q^2=m_U^2 = m_D^2=m_{3/2}^2
,\nonumber\\ m_{H_2}^2&=& m_{3/2}^2 (1- \delta)
,\label{soft1}\end{eqnarray} where values of $\delta$ of order
unity are preferred by the electroweak symmetry breaking. As we
will discuss in the next section, the SUSY corrections to the
bottom mass require $\delta$ to be close $0.3$. In this case we
find that $B$ is sufficiently small at the weak scale, which is
very important for successful electroweak breaking with such
large value of $\tan \beta$. Figure (1) shows the running of $B$
from $M_{GUT}$ to $M_{Z}$. \vskip 0.3cm
\begin{figure}[h]
\psfig{figure=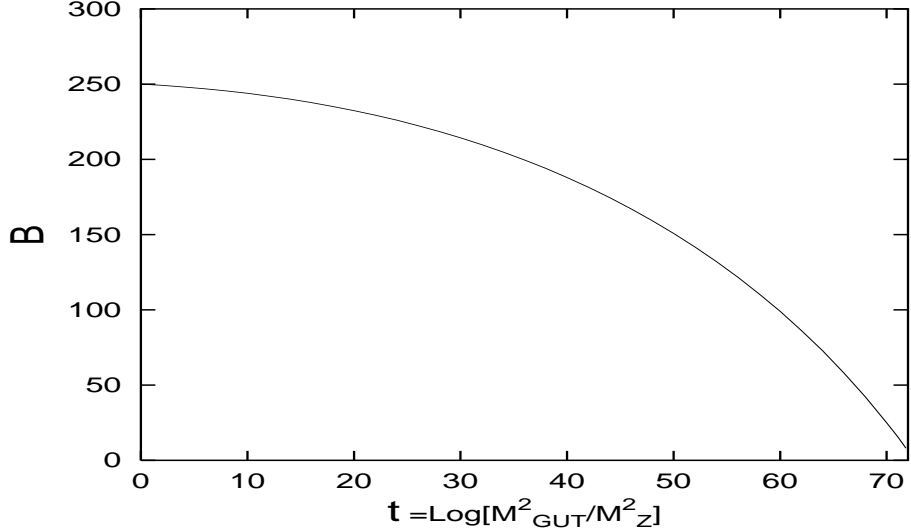,height=7cm,width=12cm} \caption{Running of the
bilinear term $B$ from $M_{GUT}$ to $M_{Z}$.}
\end{figure}
\vskip 0.3cm With the choice made in (\ref{soft1}) the Higgs
masses easily satisfy the constraint \begin{equation} m_{H_1}^2
-m_{H_2}^2 > M_Z^2 . \end{equation} Moreover, from the electroweak
breaking condition \begin{equation} \mu^2 = \frac{m_{H_1}^2
-m_{H_2}^2 \tan^2 \beta}{\tan^2 \beta -1} -M_Z^2/2 ,
\end{equation} we can determine the factor $\lambda/2\kappa$ (see
eq.(\ref{mu1})). \vskip 0.3cm
\section{SUSY correction to the bottom quark mass}
It is well known~\cite{hempfling} that in models with large $\tan
\beta$ the bottom quark mass can receive a sizable SUSY
correction. The dominant contributions are due to the
sbottom-gluino and stop-chargino loops. The tree level value of
the bottom mass is $m_b(M_Z)=\lambda_b(M_Z) v \cos \beta \simeq
3.3$ GeV, to be compared with the `measured' value~\cite{marti}
$$ m_b(M_Z) = 2.67\pm 0.50 \mathrm{GeV .}$$ We therefore would
like the SUSY corrections to be negative and $\leq 20 \% $. In
this section we estimate the SUSY corrections to $m_b$, and we
are interested in finding regions of the parameter space which
simultaneously allow small SUSY corrections and acceptable
electroweak breaking. \vskip 0.3cm The dominant contributions to
the bottom quark mass $\delta m_b$ are given by~\cite{hempfling}
\begin{equation} \delta m_b = \mu \tan \beta \left[ \frac{2
\alpha_S}{3 \pi} M_{\tilde{g}} I(m^2_{\tilde{b}_1},
m^2_{\tilde{b}_2},M_{\tilde{g}}) + \frac{\lambda_t^2}{16 \pi^2}
A_t I(m^2_{\tilde{t}_1}, m^2_{\tilde{t}_2},\mu^2) \right] ,
\label{correction}
\end{equation}
where $\lambda_t$ is the top Yukawa coupling and $$I(x,y,z)=
-\frac{xy\ln(x/y) + yz \ln(y/z) + zx \ln(z/x)}{(x-y)(y-z)(z-x)}.$$
The sign of $\delta m_b$ is the same as the sign of $\mu$. Since
we require a negative SUSY correction to reduce the tree level
value ($\simeq 3.3$ GeV) of $m_b$, we must choose $\mu < 0$. The
first contribution to $\delta m_b$ in eq.(\ref{correction}) is
the dominant one. For SUSY corrections to remain small ($\leq 20
\%$), the gluino mass and $\mu$ should not be too large. The
experimental lower limit on the gluino mass is about 150 GeV, and
so a plausible solution is to have $\mu$ small. In fact this can
be achieved for $\delta \leq 0.3$. It is important to mention that
for $\delta \leq 0$, the EW breaking conditions are not satisfied.
In Figure (2) we see that $\mu \ll m_{\tilde{q}}$ for $\delta \leq
0.3$ ($\mu \simeq 50$ GeV corresponds to $m_{\tilde{q}}\simeq
300$ GeV). We also find that the corresponding values of $\delta
m_b$ for this region of the parameter space are less than $20 \%
$. \vskip 0.3cm
\begin{figure}[h]
\psfig{figure=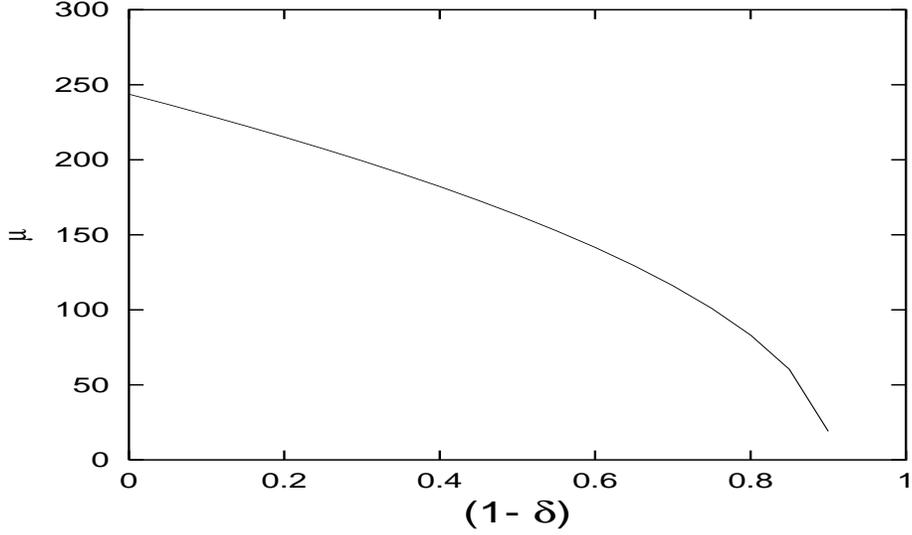,height=7cm,width=12cm} \caption{The value of
$\mu$ versus the parameter $(1-\delta)$}
\end{figure}
\vskip 0.3cm
\section{SUSY spectrum and the LSP}
In this section we investigate the SUSY spectrum in this class of
large $\tan \beta$ models arising from the parameter space that
also lead to successful EW breaking and small SUSY correction to
the mass of the bottom quark. As mentioned above, non-universality
between the gaugino masses is preferred for EW breaking and other
phenomenological aspects. \vskip0.3cm From the correction to
$m_b$ we have the constraint that the gluino mass should be
comparable (more or less) to the experimental limit, and $\mu$
should be small. This, as we will see, has important implications
for phenomenology and cosmology of these models. We observe that
with $M_1 < M_2 < M_3$ at $M_{GUT}$, the value of $M_1$ is quite
low due to the constraint on $M_3$. This implies masses for the
lightest neutralino (LSP) which are far below the experimental
limit. To avoid this we must consider sufficiently large values
for the gaugino masses (and hence $m_{3/2}$ too). Note that $M_2$
is constrained to be small from the running of the bilinear term
$B$, or else we need very large values of the trilinear coupling
to reduce the value of $B$ at $M_Z$. However, the experimental
limit on the lightest chargino can impose a lower bound on the
value of $M_2$. The gaugino mass $M_1$ is essential
unconstrained. An interesting and viable region is given by $M_3 <
M_2 < M_1$ at $M_{GUT}$. However, taking account of the different
`running', this again leads to $M_1 < M_2 < M_3$ at the weak
scale $M_Z$. \vskip 0.3cm In all of the above mentioned regions,
and for $m_{3/2}$ not too large, the LSP is expected to be
Higgsino like since $\mu$ is small. For large values of
$m_{3/2}$, $\mu$ becomes larger than $M_1$, and hence the LSP
starts to be more bino like. Figure (3) shows the neutralino
composition function $f_g$ versus the neutralino mass, with
\begin{equation}
f_g = \vert N_{11}\vert^2  + \vert N_{12} \vert^2 ,
\end{equation}
where $N_{ij}$ is the unitary matrix that diagonalize the
neutralino mass matrix. It relates the neutralino field
$\chi^0_1$ to the original ones, namely
\begin{equation}
\chi^0_1 = N_{11} \tilde{B} + N_{12} \tilde{W}^3 + N_{13}
\tilde{H}_1^0 + N_{14} \tilde{H}_2^0 .
\end{equation}
\vskip 0.3cm
\begin{figure}[h]
\psfig{figure=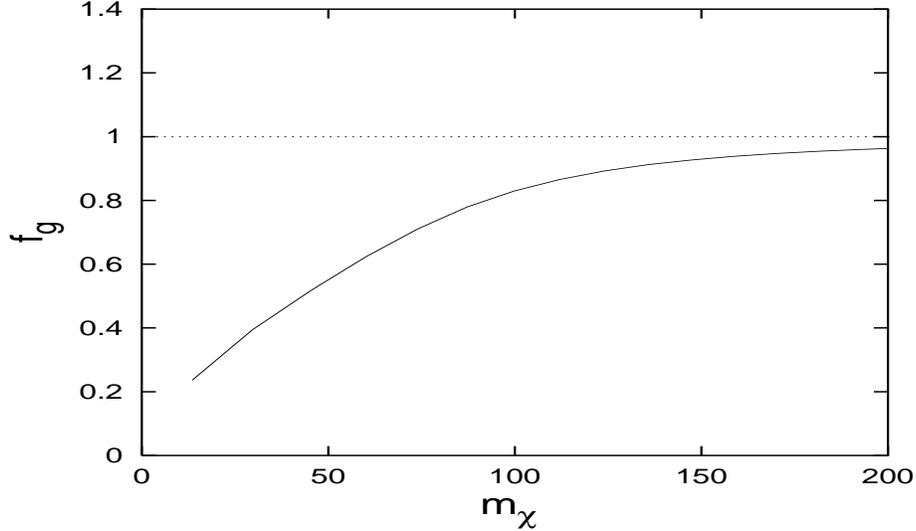,height=7cm,width=12cm} \caption{The value of
the neutralino composition function $f_g$ versus the neutralino
mass $m_{\chi}$.}
\end{figure}
\vskip 0.3cm The model is characterized by heavy SUSY scalar
masses and `light' gaugino masses. For instance, when the chargino
mass is of order the experimental bound, $m_{\chi^+} \simeq 90$
GeV, the lightest scalar mass which corresponds to one of the stop
squarks is $\simeq 500$ GeV. Furthermore, the right selectron
turns out to be the lightest slepton, with a lower bound $\sim
500$ GeV. The positivity of the eigenvalues of the stau mass
squared matrix is an important condition and usually imposes a
severe constraint on models with large $\tan \beta$
~\cite{kobayashi}. The relevant matrix is
\begin{equation}
 \left(\begin{array}{cc}
m^2_{\tilde{\tau}_L} + M_Z^2 \cos 2\beta(-\frac{1}{2} + \sin^2
\theta_W) & v Y_{\tau} (A_{\tau} \cos \beta - \mu \sin \beta)
\\
v Y_{\tau} (A_{\tau} \cos \beta - \mu \sin \beta)&
m^2_{\tilde{\tau}_R} - M_Z^2 \cos 2\beta \sin^2
\theta_W
\end{array}
\right),
\end{equation}
where $v^2=\langle H_2 \rangle^2 + \langle H_1 \rangle^2$. In the
case of large $\tan \beta$ the tau Yukawa coupling is large and
hence the off diagonal elements relative to the diagonal elements
cannot be ignored. This could lead to a negative eigenvalue.
However, with $m^2_{\tilde{\tau}_L}$ and $m^2_{\tilde{\tau}_R}$
of order $m_{3/2}^2$ at $M_{GUT}$, it turns out that this is not
the case and even the lowest eigenvalue of this matrix is larger
than the mass squared of the right selectron. \vskip 0.3cm
\section{Higgs spectrum} In the limit $\lambda=\kappa$, the
superpotential $W= \kappa S [ (H\bar{H} -M^2) + h^2]$ has an
accidental $U(4)$ symmetry under which $(H, h^{(1)})\in 4$ and
$(\bar{H}, \varepsilon h^{(2)})\in \bar{4}$, \ie , they transform
as the fundamental and antifundamental representation
respectively. When $H$ and $\bar{H}$ acquire their VEV, the
$U(4)$ symmetry breaks to $U(3)$. Hence, we expects seven
`goldstone' superfields, only three of which are true goldstone
superfields that are absorbed by the massive gauge superfields.
The remaining four superfields correspond to the physical state
$h^{(1)}$ and $h^{(2)}$. This accidental symmetry of the
superpotential is broken when supersymmetry is broken, so that
$h^{(1)}$ and $h^{(2)}$ are `psuedogoldstone' bosons. For
$\lambda \neq \kappa$ the $U(4)$ symmetry is explicitly broken in
the superpotential and the above arguments must be reconsidered.
\vskip 0.3cm The lightest Higgs scalar $(h^0)$ has the well-known
mass at tree level,
\begin{equation}
m^2_{h^0} = \frac{1}{2} (m_A^2 +m_Z^2 - \sqrt{(m_A^2 + m_Z^2)^2 -
4m_Z^2 m_A^2 \cos^2 2\beta}), \label{mh0}
\end{equation}
where \begin{equation} m_A^2 = m_{H_1}^2 + m_{H_2}^2 + 2 \mu^2 .
\label{ma} \end{equation} For $\tan \beta \simeq \frac{m_t}{m_b}$,
eq.(\ref{mh0}) gives $$ m_{h^0} \simeq m_Z .$$ However, one loop
corrections~\cite{haber1} can increase this by about (40-60) GeV,
while two loop corrections~\cite{haber2} can lower the value by
approximately 10 GeV (see also~\cite{shafi3}). \vskip 0.3cm Since
the value of $\mu$ is quite constrained, we expect that the
neutral pesudoscalar Higgs boson $A$, whose mass $m_A$ is given
in eq.(\ref{ma}), is not too heavy. Indeed we find that $m_A
\simeq 100$ GeV for $m_{3/2}\simeq 500$ GeV. However, $h^0$ turns
out to be the lightest supersymmetric Higgs. In Figure 4 we
display the correlation between the masses of $h^0$ and $A$.
\vskip 0.3cm
\begin{figure}[h]
\psfig{figure=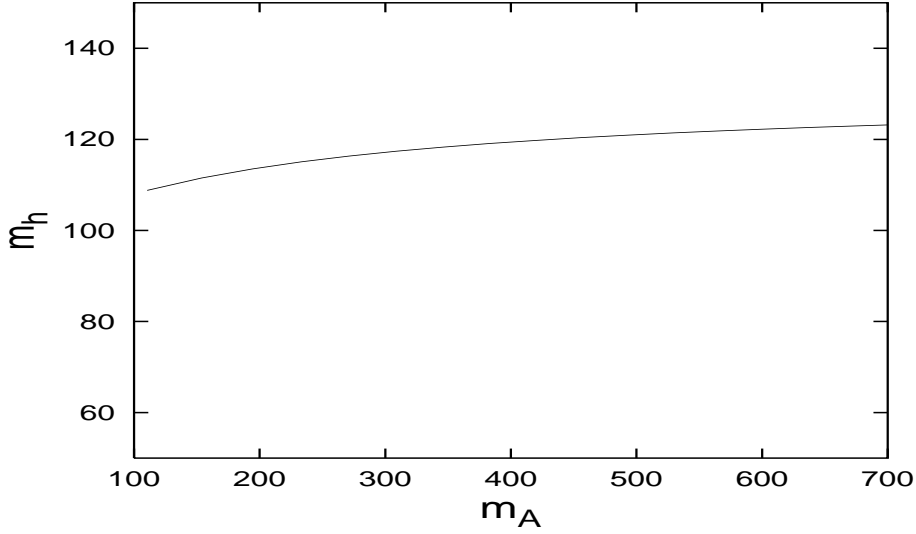,height=7cm,width=12cm} \caption{Correlation
of masses of the lightest and psuedoscalar Higgses}
\end{figure}
\vskip 0.3cm One could expect that in the region where $m_A$ is of
order $\mathcal{O}(100)$ GeV, the charged Higgs boson mass is of
the same magnitude,  $$ m_{H^{\pm}}^2 = m_Z^2 + m_A^2 , $$ which
may lead to a large value for the branching ratio of $ b
\rightarrow s \gamma$. However, the chargino mass in this model
is very close to the experimental bound, so that we have a large
chargino contribution which gives rise to destructive
interference with the SM and charged Higgs contributions, as will
be explained in the next section. Hence, a relatively light
psuedoscalar Higgs is allowed.
\section{Constraints from $b \rightarrow s \gamma$}
In this section we focus on the constraints on the parameter space
which arise from the decay $b \rightarrow s \gamma$. The CLEO
experiment \cite{amer} has confirmed that $ 1\times 10^{-4} <
BR(b \rightarrow s \gamma) < 4 \times 10^{-4}$. In supersymmetric
models there are three significant contributions to the total
amplitude, namely from the $W$-loop, charged Higgs loop and the
chargino loop. The inclusive branching ratio for $b \rightarrow s
\gamma$ is given by
\begin{equation}
R = \frac{BR(b \rightarrow s \gamma)}{BR(b \rightarrow c
e\bar{\nu})}. \end{equation} The computation of $R$
yields~\cite{bs}
\begin{equation}
R= \frac{\mid V_{ts}^* V_{tb}\mid^2}{\mid V_{cb} \mid^2} \frac{6
\alpha_{em}}{\pi} \frac{[\eta^{16/23} A_{\gamma} + \frac{8}{3}
(\eta^{14/23} -\eta^{16/23}) A_g +
C]^2}{I(x_{cb})[1-\frac{2}{3\pi} \alpha_S(m_b) f(x_{cb})]}.
\end{equation} Here, $\eta =
\frac{\alpha_S(m_W)}{\alpha_S(m_b)}$, and  $C$ represents the
leading-order QCD corrections to $b \rightarrow s \gamma$
amplitude at the scale $Q=m_b$~\cite{misiak2}. The function
$I(x)$ is given by $$I(x)=1-8 x^2 +8 x^6 -x^8 -24 x^4 \ln x, $$
where $x_{cb} = \frac{m_c}{m_b}$, and $f(x)$ is a QCD correction
factor, with $ f(x_{cb})=2.41$. The amplitude $A_{\gamma}$ is from
the photon penguin vertex, the amplitude $A_g$ is from the gluon
penguin vertex, and they are given in Ref.~\cite{bs}. It was
shown in MSSM~\cite{bs}, and in models with dilaton dominated SUSY
breaking~\cite{khalil} that with $\tan \beta \simeq 2$, the
chargino contribution gives rise to a destructive interference
with SM contribution and charged Higgs ($H^+$) contribution, but
it is generally smaller than the latter. This leads to a severe
constraint on the parameter space of these models. It was also
realized that the constraint is less severe if the soft terms are
non-universal. In the moduli-dominant SUSY breaking
model~\cite{khalil2} it was shown that the chargino contribution
gives rise to substantial destructive interference with SM and
$H^+$ amplitude, so that the branching ratio of $b \rightarrow s
\gamma$ is less than the SM value. \vskip 0.3cm In figure (5) we
show that we have a similar situation here since the model is
characterized by 'not too large' guagino masses. Therefore, the
chargino contribution which is inversely proportional to its mass
square becomes significant. This result is quite interesting
since, as pointed out in~\cite{misiak}, the SM prediction is
above the CLEO measurement at the $1 \sigma$ level. Hence, any new
physics beyond the SM should provide destructive interference
with the SM amplitude and our model has this feature. \vskip 0.3cm
\begin{figure}[h]
\psfig{figure=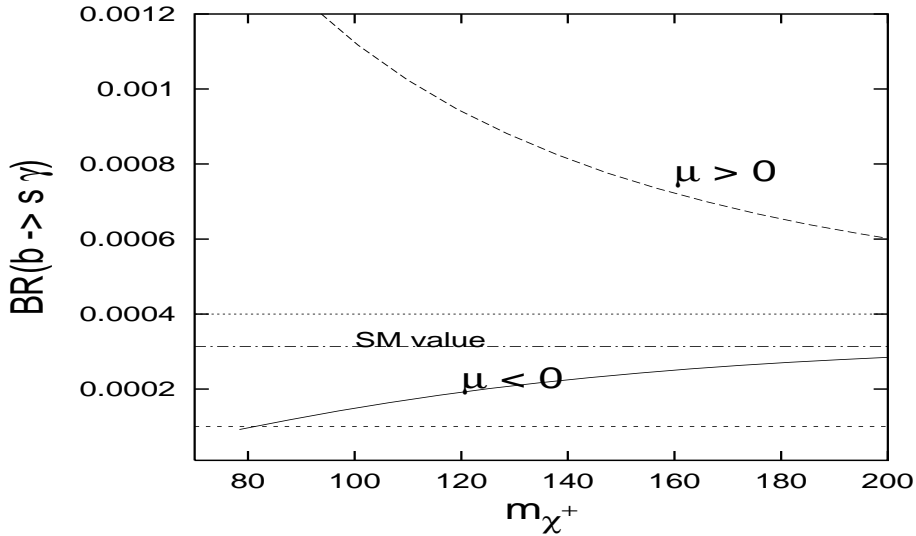,height=7cm,width=12cm} \caption{The branching
ratio of $b\rightarrow s \gamma$ versus the lightest chargino
mass. The horizontal lines at $1\times 10^{-4}$ and $4\times
10^{-4}$ correspond to the experimental bounds.}
\end{figure}
\vskip 0.3cm It is important to note that for $m_{\chi^+}\simeq
200$ GeV the gravitino mass $m_{3/2}$ is of order 1.5 TeV.
Furthermore, for this value of $m_{3/2}$ the signs of the quantity
$M_2^2 -\mu^2 -2 M_W^2 \cos 2\beta$ as well as the chargino
contribution are reversed. Thus, destructive interference from
the chargino contribution now corresponds to the case $\mu
>0$ (instead of $\mu <0$). However, from $\delta m_b < 0$, we
obtained the constraint that $\mu$ should be negative. Therefore,
from the supersymmetric correction to the bottom quark mass and
the experimental bound on the branching ratio of the
$b\rightarrow s \gamma$ we find an upper bound on $m_{3/2}$ of
about 1.5 TeV. \vskip 0.35cm
\section{Enhancement of $\mu \rightarrow e \gamma$ at large $\tan
\beta$} Lepton flavor violation (LFV) is considered a significant
prediction of many supersymmetric models and provides a sensitive
probe of physics beyond the standard model. In this section we
show that with $\tan \beta \simeq m_t/m_b$ the LFV process $\mu
\rightarrow e \gamma$ may be enhanced in the class of models under
discussion and presumably amenable to ongoing and planned
experminets.\vskip 0.3cm This result can be understood as follows.
After symmetry breaking we can write the superpotential in
(\ref{susypot}) as
\begin{equation}
W= W_{MSSM} + \lambda_{\nu i} V_{\nu CKM}^{ij} \bar{N}_i L_j H_2 +
M_{\nu R}^{ij} \bar{N}_i \bar{N}_j .
\end{equation}
Here, $i,j =1,...,3$ are the generation indices, and the
superfields $L$ and $\bar{N}$ represent the leptons $(\nu_L, e_L)$
and $\nu_R^c$ respectively. The lepton sector has a mixing matrix
$V_{\nu CKM}$ analogous to the CKM matrix in the quark sector,
which contributes to lepton flavor violation. In particular, the
off-diagonal components of the matrices $m_{\tilde{e}}^2$,
$m_{\tilde{l}}^2$ and $A_l$ are the sources for LFV.\\ \vskip
0.3cm In our model, the amplitude of the photino contribution is
given in terms of the mass insertion $\delta^l_{AB}$ defined by
$\delta^l_{AB} = \frac{\Delta^l_{AB}}{\tilde{m}^2}$, where
$\tilde{m}$ is an average slepton mass and $\Delta^l$ denote the
off-diagonal terms in the slepton mass matrices. The mass
insertion to accomplish the transition from $\tilde{l}_{i}$ to
$\tilde{l}_{j}$ is given by~\cite{hisano}
\begin{eqnarray}
(\Delta^l_{LL})_{ij}&\simeq&-\frac{1}{8\pi^2} \lambda^2_{\nu 3}
(V_{\nu CKM}^{\dag})^{i3} V_{\nu CKM}^{3j} (3 m_{3/2}^2 + A^2)
\ln(\frac{M_{GUT}}{M_{\nu R}}) ,\label{delta1}
\\
(\Delta^l_{LR})_{ij}&\simeq&-\frac{1}{8\pi^2}  \lambda^2_{\nu 3}
(V_{\nu CKM}^{\dag})^{i3} V_{\nu CKM}^{3j}  \lambda_{l j} v A
\ln(\frac{M_{GUT}}{M_{\nu R}}) ,
\\
(\Delta^l_{RL})_{ij}&=& (\Delta^l_{LR})^{\dag}_{ij} ,
\\
(\Delta^d_{RR})_{ij}&=&0.\label{delta4}
\end{eqnarray}
Here the neutrino Yukawa coupling constants except for
$\lambda_{\nu 3}$ are ignored. Since $\Delta^l_{LR}$ is
proportional to $\lambda_l v = m_l/\cos\beta$ this quantity, and
hence the branching ratio, is enhanced for large $\tan \beta$.
The branching ratio for the process $\mu \rightarrow e \gamma$ is
given by~\cite{masiero}
\begin{equation}
BR(\mu \rightarrow e \gamma) = \frac{\alpha^3}{G_F^2}
\frac{12\pi}{m_{\tilde{l}}^4} \left[ \vert M_3(x)
(\delta^l_{21})_{LL} + \frac{m_{\tilde{\gamma}}}{m_l} M_1(x)
(\delta^l_{21})_{LR}\vert^2 + L \leftrightarrow R \right] BR(\mu
\rightarrow e \nu \bar{\nu}),
\end{equation}
where $x=\frac{m_{\tilde{\gamma}}}{m_{\tilde{l}}}$, and the
functions $M_1(x)$ and $M_3(x)$ are given by
\begin{eqnarray}
M_1(x)&=& \frac{1+4 x - 5 x^2 + 4 x \ln(x) + 2 x^2
\ln(x)}{2(1-x)^4} ,\\ M_3(x)&=& \frac{-1+9 x + 9 x^2 -17 x^3 + 18
x^2 \ln(x) + 6 x^3 \ln(x)}{12(x-1)^5}.
\end{eqnarray}

From eqs.(\ref{delta1}-\ref{delta4}) the branching ratio depends
on the neutrino Yukawa couplings. Several forms for these Yukawa
matrices were studied in the supersymmetric $SU(4)_C\times SU(2)_L
\times SU(2)_R$ model~\cite{king2}. Here we consider the ansatz
given in Ref.\cite{pati} which is compatible with the solar and
the atmospheric neutrino data. It remains to be seen whether this
ansatz or some form close to it can be realized in the present
scheme which contains $U(1)_R$ symmetry. \vskip 0.3cm In Figure
(6) we exhibit the branching ratio $BR(\mu \rightarrow e \gamma)$
versus the chargino mass. It is interesting that the predicted
values of the branching ratio are very close to the experimental
bound, and even for very heavy sleptons only about one order of
magnitude below the current limits.  \vskip 0.3cm
\begin{figure}[h]
\psfig{figure=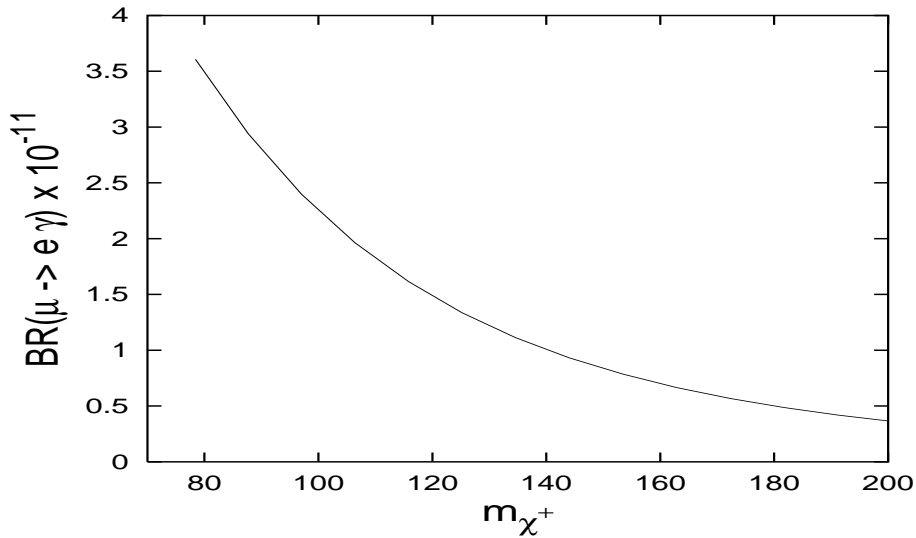,height=7cm,width=12cm} \caption{The
branching ratio $BR(\mu \rightarrow e \gamma)$ versus the lightest
chargino mass}
\end{figure}
\vskip 0.3cm
\section{conclusions}
We have studied the low energy consequences of a class of
supersymmetric models with left-right symmetry, in particular the
$SU(4)_C\times SU(2)_L\times SU(2)_R$ scheme. In these models the
gauge hierarchy and $\mu$ problems are first resolved and $\tan
\beta$ is constrained to be of order $m_t/m_b$. We have shown
that non-universality between $m^2_{H_1}$ and $m^2_{H_2}$ is
favorable for successful EW symmetry breaking. On the other hand,
the requirement that SUSY corrections to the bottom quark mass
should not be exceed $20 \% $ gives strong constraints on the
allowed parameter space, namely it leads to $\mu <0$ and not too
large, while the gluino mass should be of order the experimental
bound. \vskip 0.3cm We have investigated the SUSY spectrum in this
class of large $\tan \beta$ models. It turns out that the lightest
chargino and neutralino are almost gaugino-like for large ($\sim
TeV$) values of $m_{3/2}$, and they become more Higgsino-like if
$m_{3/2}$ is not too large, since in this region $\mu$ is small.
Furthermore, we have shown that the lightest Higgs mass is of
order 120 GeV, and the neutral pesudoscalar Higgs boson $A$ is
not too heavy($\sim \mathcal{O}(100)$ GeV).\vskip 0.3cm We also
examined the radiative process $b\rightarrow s\gamma$. This
process imposes the constraint that the gravitino mass to be less
than 1.5 TeV. Finally, we find that the LFV process $\mu
\rightarrow e \gamma$ is expected to be enhanced due to the large
value of $\tan\beta$.

\section*{Acknowledgments}
We would like to thank G.Dvali for useful discussions. S.K. would
like to acknowledge the financial support provided by the
Fulbright Commission and the hospitality of the Bartol Research
Institute. Q.S. is supported in part by DOE Grant No.\
DE-FG02-91ER40626, and by Nato contract number CRG-970149.

\providecommand{\bysame}{\leavevmode\hbox
to3em{\hrulefill}\thinspace}

\end{document}